\documentclass[11pt, oneside, reqno]{amsart}
\usepackage[utf8]{inputenc}

\usepackage{amsmath, amsfonts, amssymb, amsthm}
\usepackage{physics, enumerate, dsfont}

\newtheorem{theorem}{Theorem}[section]

\newtheorem{corollary}[theorem]{Corollary}

\newtheorem{remark}[theorem]{Remark}

\renewcommand{\Re}{\operatorname{Re}}

\begin{document}
\title[Excess charge for bosonic systems with Coulomb potentials]{\textbf{Bound on the excess charge for bosonic systems interacting through Coulomb potentials}}

\author[RB]{Rafael D. Benguria}
\author[JMG]{Juan Manuel Gonz\'alez--Brantes}

\address{Instituto de F\'isica, Av. Vicu\~na Mackenna 4860, Santiago, Chile.}
\email{rbenguri@fis.puc.cl}
\address{Facultad de F\'isica, Av. Vicu\~na Mackenna 4860, Santiago, Chile.}
\email{jmgonzalez4@uc.cl}

\begin{abstract} 
We prove an upper bound on the excess charge for non-relativistic atomic systems, independent of the particle statistics. This result improves Lieb's bound of 1984 for any $Z \ge 12$.
\end{abstract}

\maketitle

\vspace*{-0.5 cm}
\noindent MSC2020--Mathematics Subject Classification System: 81V73, 81V45, 35A15, 35A01. 

\section{Introduction}
The goal of this article is to prove the following upper bound on the excess charge, $N - Z$, of an atom.

\begin{theorem}[Bound on the excess charge for bosonic atoms] 
\label{thm:excess_charge} 
Let us consider the $N$-particle Hamiltonian

\begin{equation}
\label{eq:many_body_hamiltonian}
H_{N, Z} = \sum_{j = 1}^{N} \qty(-\Delta_{j} - \dfrac{Z}{\abs{x_{j}}}) + \sum_{1 \leq i < j \leq N} \dfrac{1}{\abs{x_{i} - x_{j}}},
\end{equation}

\noindent acting on the symmetric tensor product of $N$ copies of $L^{2}\qty(\mathds{R}^{3})$ (Bosonic Atom). Then, the maximum number of particles, $N$, for which the Hamiltonian has a bound state satisfies

\begin{equation}
\label{eq:excess_charge}
N < 1.4811Z + 3.1516Z^{1/3},
\end{equation}

\noindent for $Z \geq 12$.
\end{theorem}

\begin{remark} This upper bound improves Lieb's result \cite{Lieb1984-1}, $N(Z)<2Z+1$, for any $Z \ge 12$ in the bosonic case. 
\end{remark}

We first recall some ideas and results from \cite{BenguriaTubino2022}. In their paper, and using an scaling argument, the authors establish a Virial Theorem on the energy components $K$, $A$, and $R$ of the \emph{one-particle Hartree atomic functional} which is defined by

\begin{equation}
\label{eq:1_particle_hartree_atomic_functional}
\mathcal{E}(\psi) = \underbrace{\int_{\mathds{R}^{3}} \abs{\nabla \psi}^{2} \dd{x}}_{=:K} - \underbrace{\int_{\mathds{R}^{3}} \dfrac{Z}{\abs{x}} \psi^{2} \dd{x}}_{=:A} + \underbrace{\dfrac{1}{2} \int_{\mathds{R}^{3}} \int_{\mathds{R}^{3}} \psi^{2}\qty(x) \dfrac{1}{\abs{x - x'}} \psi^{2}\qty(x') \dd{x'} \dd{x}}_{=:R},
\end{equation}

\noindent to use later the \emph{Coulomb Uncertainty Principle},

\begin{equation}
\label{eq:coulomb_uncertainty_principle}
\int_{\mathds{R}^{3}} \dfrac{1}{\abs{x}} \psi\qty(x)^{2} \dd{x} \leq \norm{\nabla \psi}_{2} \norm{\psi}_{2},
\end{equation}

\noindent to get bounds on the energy components $K$, $A$, and $R$. Results about existence and uniqueness of minimizers of \eqref{eq:1_particle_hartree_atomic_functional} can be also found in \cite{BenguriaTubino2022} and references therein. In this way, to establish the analytic bound on the excess charge for the one-particle Hartree model

\begin{equation}
\label{eq:benguria_tubino_bound}
N \leq \dfrac{5}{4\beta}Z \approx 1.5211Z,
\end{equation}

\noindent the authors followed the strategy developed by P.~T.~Nam in \cite{Nam2012} on the associated \emph{Hartree equation},

\begin{equation}
	\label{eq:hartree_equation}
	-\Delta \psi = \phi(x)\psi,
\end{equation}

\noindent with

\begin{equation}
\label{eq:coulomb_potential}	
\phi(x) = \frac{Z}{|x|} - \int_{\mathds{R}^{3}} \frac{\psi^{2}(x')}{|x-x'|}\dd{x'}
\end{equation}

\noindent the Coulomb potential. The quantity $\beta$ in \eqref{eq:benguria_tubino_bound} is also defined in \cite{Nam2012}. To make this paper self-contained, we revisit this strategy. Let us multiply \eqref{eq:hartree_equation} by $\abs{x}^{2} \ \psi\qty(x)$ and integrate all over $\mathds{R}^{3}$,

\begin{equation}
\label{eq:nam_strategy_1}
\qty(\abs{x}^{2}\psi\qty(x), -\Delta \psi\qty(x)) = \qty(\abs{x}^{2}\phi\qty(x), \psi^{2}\qty(x)).
\end{equation}

\noindent The inner product on the left side of \eqref{eq:nam_strategy_1} can be bound by,

\begin{equation}
\label{eq:nam_strategy_2}
\qty(\abs{x}^{2}\psi\qty(x), -\Delta \psi\qty(x)) \geq -\dfrac{3}{4}N,
\end{equation}

\noindent see, e.g., \cite{Nam2012}. Here,

\begin{equation}
	\label{eq:number_particles}
	N = \int_{\mathds{R}^{3}} \psi(x)^{2} \dd{x}.
\end{equation}

\noindent Using \eqref{eq:coulomb_potential}, we then have for the right side of \eqref{eq:nam_strategy_1} that

\begin{equation}
\label{eq:nam_strategy_3}
Z\qty(\psi\qty(x), \abs{x}\psi\qty(x)) - \int_{\mathds{R}^{3}} \int_{\mathds{R}^{3}} \psi^{2}\qty(x) \dfrac{\abs{x}^{2}}{\abs{x - x'}} \psi^{2}\qty(x') \dd{x'} \dd{x} \geq -\dfrac{3}{4}N.
\end{equation}

\noindent After applying standard symmetrization on the second term on the left side of \eqref{eq:nam_strategy_3}, we use the definition of $\beta$

\begin{equation}
\label{eq:beta_definition}
\beta := \inf_{\psi} \dfrac{1}{2} \dfrac{\displaystyle{\int_{\mathds{R}^{3}} \int_{\mathds{R}^{3}} \psi^{2}(x) \dfrac{\abs{x}^{2} + \abs{x'}^{2}}{\abs{x - x'}} \psi^{2}\qty(x') \dd{x'} \dd{x}}}{\displaystyle{\qty(\int_{\mathds{R}^{3}} \abs{x} \psi^{2}\qty(x) \dd{x}) \qty(\int_{\mathds{R}^{3}} \psi^{2}\qty(x) \dd{x})}},
\end{equation}

\noindent where the minimum is taken overall $\psi \in L^{2}(\mathds{R}^{3})$. Here, $\beta \in [0.8218, 0.8705)$ \cite{Nam2012}. It follows from \eqref{eq:nam_strategy_3} and \eqref{eq:beta_definition} that

\begin{equation}
\label{eq:nam_strategy_4}
Z\qty(\psi\qty(x), \abs{x}\psi\qty(x)) - \beta N \qty(\psi\qty(x), \abs{x}\psi\qty(x)) \geq -\dfrac{3}{4}N.
\end{equation}

Notice that the quantity $\qty(\psi\qty(x), \abs{x}\psi\qty(x))$, defined as $J$ in \cite{BenguriaTubino2022}, can be understood as the average distance of one particle with respect to the nucleus. Its $N$-particle counterpart, $I$, will be relevant below. Using an appropriate Cauchy--Schwarz inequality to write $N^{2}$ (see, e.g., \cite{BenguriaTubino2022, GonzalezBrantes2024}) one has that

\begin{equation}
\label{eq:nam_strategy_5}
J \geq 3\dfrac{N}{Z}.
\end{equation}

\noindent Finally, \eqref{eq:benguria_tubino_bound} follows from \eqref{eq:nam_strategy_4} and \eqref{eq:nam_strategy_5}.

\medskip

Eventually, we shall also consider the following corollary, which is a consequence from the Virial Theorem on $K$, $A$, and $R$ stated in \cite{BenguriaTubino2022, GonzalezBrantes2024}.

\begin{corollary}[Lower bound on $\mathcal{E}\qty(\psi)$] 
\label{thm:benguria_jm_energybound}
\noindent Let $\mathcal{E}\qty(\psi)$ be the one-particle Hartree atomic functional, as defined by \eqref{eq:1_particle_hartree_atomic_functional}. For any $\psi \in H^{1}\qty(\mathds{R}^{3})$, we have the following lower bound,

\begin{equation}
\label{eq:benguria_jm_energybound}
\mathcal{E}\qty[\psi] \geq -k Z^{3},
\end{equation}

\noindent with $k = N_{\textrm{c}}/9Z \approx 0.1344$. Here, $N_{\textrm{c}}$ is the largest value of \eqref{eq:number_particles} for which there is a minimizer of $\mathcal{E}(\psi)$ in $H^{1}(\mathds{R}^{3})$. This minimizer is the unique positive solution of the Hartree equation \eqref{eq:hartree_equation}. Numerically, $N_{\textrm{c}} \approx 1.21Z$ (see \cite{Baumgartner1984}).
\end{corollary}

\begin{remark} It follows from the Virial Theorem for the Hartree equation \cite{BenguriaTubino2022, GonzalezBrantes2024} that for $N \leq N_{\textrm{c}}$, one has
	
\begin{equation}
\label{}
E = -K = -\frac{A}{3} \geq -\frac{1}{9}NZ^{2}.	
\end{equation}

\noindent On the other hand, $E(N)$ is a monotone decreasing function of $N$ for $N \leq N_{\textrm{c}}$ and $E(N) = E(N_{\textrm{c}})$ for $N > N_{\textrm{c}}$. For a proof of this fact see, e.g., \cite{BBL1981, Lieb1981}. In both cases, take $p=5/3$ and $\gamma=0$. Therefore,

\begin{equation}
\label{}
E(N) \geq -\frac{1}{9}N_{\textrm{c}}(Z)Z^{2} = -\frac{1}{9}\frac{N_{c}(Z)}{Z}Z^{3} = -kZ^{3},
\end{equation}

\noindent where $N_{\textrm{c}}(Z)/Z$ stands for Baumgartner's value \cite{Baumgartner1984}.

\end{remark}

\begin{remark}
In \cite{GonzalezBrantes2024} there is a numerical improvement on $k = N_{\textrm{c}}/9Z \approx 0.1344$ to $C_{\emph{IE}} \approx 0.1242$.
\end{remark}

We are now able to study the $N$-particle Hartree atomic case. Here, the $N$-particle Hartree atomic functional is given by,

\begin{align}
\label{eq:manybody_hartree_atomic_expectation}
\left \langle \psi_{N, Z}, H_{N, Z}\psi_{N, Z} \right \rangle &= K_{\psi} - \left \langle \psi_{N, Z}, \sum_{j = 1}^{N} \dfrac{Z}{\abs{x_{j}}} \psi_{N, Z} \right \rangle + \left \langle \psi_{N, Z}, \sum_{1 \leq i < j \leq N} \dfrac{1}{\abs{x_{i} - x_{j}}} \psi_{N, Z} \right \rangle, 
\end{align}

\noindent with $H_{N, Z}$ defined by \eqref{eq:many_body_hamiltonian} and $K_{\psi}$ by Theorem \ref{thm:hoffmann_ostenhof_inequality} below. To do so, we consider a series of well-known results concerning the behavior of $N$-particle systems describing Quantum Mechanics. Before going to these results, let us define

\begin{equation}
\label{SPD}
\rho_{\psi}\qty(x) = N \int_{\mathds{R}^{3\qty(N - 1)}} \abs{\psi\qty(x, x_{2}, \dots, x_{N})}^{2} \dd{x_{2}} \dots \dd{x_{N}},
\end{equation}

\noindent to be the \emph{Single Particle Density} associated to the wave-function $\psi_{N, Z} \in H^{1}\qty(\mathds{R}^{3N})$. 

\begin{theorem}[Kinetic Energy Bound in terms of the density; Maria and Thomas Hoffmann-Ostenhof, 1977 \cite{HoffmannOstenhof1977}]
\label{thm:hoffmann_ostenhof_inequality} 
For any $N$-particle wave-function $\psi_{N, Z} \in H^{1}\qty(\mathds{R}^{3N})$ one has

\begin{equation}
\label{eq:hoffmann_ostenhof_inequality}
K_{\psi} := \left \langle \psi_{N, Z}, \sum_{i = 1}^{N} -\Delta_{i} \psi_{N, Z} \right \rangle \geq \qty(\sqrt{\rho_{\psi}}, -\Delta \sqrt{\rho_{\psi}}) =: K_{\rho_{\psi}}.
\end{equation}

\end{theorem}

\noindent Using Theorem \ref{thm:hoffmann_ostenhof_inequality}, we can bound the $N$-particle kinetic energy. See, e.g., \cite{GonzalezBrantes2024} and remarks on Hoffmann-Ostenhof inequality therein. The following theorem establishes the also called \emph{exchange and correlation} inequality.

\begin{theorem}[Bound on the Coulomb Energy in terms of the density; E.~H.~Lieb and S.~Oxford, 1981 \cite{LiebOxford1981}] \label{thm:lieb_oxford_inequality} For any $N$-particle wave-function $\psi_{N, Z} \in H^{1}\qty(\mathds{R}^{3N})$, with density given by (\ref{SPD}), one has

\begin{align}
\label{eq:lieb_oxford_inequality}
\left \langle \psi_{N, Z}, \sum_{1 \leq i < j \leq N} \dfrac{1}{\abs{x_{i} - x_{j}}} \psi_{N, Z} \right \rangle &\geq \dfrac{1}{2} \int_{\mathds{R}^{3}} \int_{\mathds{R}^{3}} \dfrac{\rho_{\psi}\qty(x) \rho_{\psi}\qty(x')}{\abs{x - x'}} \dd{x'} \dd{x} - C_{\emph{LO}} \int_{\mathds{R}^{3}} \rho_{\psi}\qty(x)^{4/3} \dd{x}.
\end{align}

\end{theorem}

\noindent $C_{\textrm{LO}}$ is the Lieb-Oxford constant which was recently improved by M. Lewin, E. Lieb, and R. Seiringer to be $1.5765$ \cite{LewinLiebSeiringer2022}. Using (\ref{SPD}), it is straighforward to show that, for any $N$-particle wave-function $\psi_{N, Z} \in H^{1}\qty(\mathds{R}^{3N})$, we have,

\begin{equation}
\label{eq:exact_attraction_term}
A_{\psi}:=\left \langle \psi_{N, Z}, \sum_{j = 1}^{N} \dfrac{Z}{\abs{x_{j}}} \psi_{N, Z} \right \rangle = \int_{\mathds{R}^{3}} \dfrac{Z}{\abs{x}} \rho_{\psi}\qty(x) \dd{x}.
\end{equation}

\noindent At the right side of the inequality given by \eqref{eq:lieb_oxford_inequality}, appears an error term with the Lieb-Oxford constant as a prefactor. We control this term using the following inequality,

\begin{align}
\label{eq:gagliardo_inequality}
\int_{\mathds{R}^{3}} \rho_{\psi}\qty(x)^{4/3} \dd{x} &\leq C_{\textrm{GNS}} \qty(\int_{\mathds{R}^{3}} \qty(\nabla \sqrt{\rho_{\psi}})^{2} \dd{x})^{1/2} \qty(\int_{\mathds{R}^{3}} \rho_{\psi}\qty(x) \dd{x})^{5/6} = C_{\textrm{GNS}} K_{\rho_{\psi}}^{1/2} N^{5/6}.
\end{align}

\noindent This inequality is a special case of the Gagliardo-Nirenberg-Sobolev inequality (GNS inequality, for short). In \eqref{eq:gagliardo_inequality}, $C_{\textrm{GNS}}$ stands for the GNS constant. This constant was computed numerically in \cite{GonzalezBrantes2024} to be $0.2812$. See, e.g., \cite{BeVaVDB019, GonzalezBrantes2024} and all references therein about technical aspects on GNS-type inequalities.

\medskip

Putting all this together, one has that for any $N$-particle wave-function $\psi_{N, Z} \in H^{1}\qty(\mathds{R}^{3N})$ with density $\rho_{\psi}$,

\begin{equation}
\label{eq:connection_without_error_term}
\left \langle \psi_{N, Z}, H_{N, Z}\psi_{N, Z} \right \rangle \geq \mathcal{E}(\sqrt{\rho_{\psi}}) - C_{\textrm{LO}} \int_{\mathds{R}^{3}} \rho_{\psi}\qty(x)^{4/3} \dd{x}.
\end{equation}

\noindent This emphasizes the connection between the one--particle Hartree atomic model with the $N$--particle Hartree atomic model. Using \eqref{eq:gagliardo_inequality} and \eqref{eq:connection_without_error_term}, we get,

\begin{equation}
\label{eq:connection_with_error_term}
\left \langle \psi_{N, Z}, H_{N, Z}\psi_{N, Z} \right \rangle \geq \mathcal{E}(\sqrt{\rho_{\psi}}) - DK_{\rho_{\psi}}^{1/2}N^{5/6},
\end{equation}

\noindent with $D:=C_{\textrm{LO}}C_{\textrm{GNS}} \leq 1.5765 \cdot 0.2812 = 0.4433$. 

\section{Proof of Theorem \ref{thm:excess_charge}.}
\proof We shall divide this proof in four parts.

\subsection*{Estimating $E\qty(N, Z)$.} We use \eqref{eq:benguria_jm_energybound} with $k = C_{\textrm{IE}}$ on \eqref{eq:connection_with_error_term},

\begin{equation}
\label{eq:estimating_1}
E\qty(N, Z) \geq -C_{\textrm{IE}}Z^{3} - DK_{\rho_{\psi}}^{1/2}N^{5/6},
\end{equation}

\noindent the Virial Theorem in its $N$-particle version, $E\qty(N, Z) = -K_{\psi}$, and \eqref{eq:hoffmann_ostenhof_inequality}, to get

\begin{equation}
\label{eq:estimating_2}
E\qty(N, Z) \geq -C_{\textrm{IE}}Z^{3} - D\sqrt{-E\qty(N, Z)}N^{5/6}.
\end{equation}

\noindent Now, we use that $E(N,Z) \ge -NZ^{2}/2$ \cite{Lieb1976, Lieb1990} in \eqref{eq:estimating_2} to get rid of the term $\sqrt{-E(N,Z)}$ in \eqref{eq:estimating_2},

\begin{equation}
\label{eq:estimating_3}
E\qty(N, Z) \geq -C_{\textrm{IE}}Z^{3} - D\sqrt{\dfrac{1}{2}}N^{4/3}Z.
\end{equation}

\subsection*{The eigenvalue equation.}
\noindent Let us assume that $E\qty(N, Z)$ is an eigenvalue of $H_{N, Z}$ corresponding to some normalized eigenfunction $\psi_{N, Z} \in H^{1}\qty(\mathds{R}^{3N})$, i.e., 

\begin{equation}
\label{eq:eigenvalue_1}
\qty(H_{N, Z} - E\qty(N, Z)) \psi_{N, Z} = 0.
\end{equation}

\noindent Following \cite{Nam2012}, multiply \eqref{eq:eigenvalue_1} by $\abs{x_{N}}^{2} \ \overline{\psi}_{N, Z}$ and integrate,

\begin{align}
\label{eq:eigenvalue_2}
0 &= \left \langle \abs{x_{N}}^{2} \ \psi_{N, Z}, \qty(H_{N - 1, Z} - E\qty(N, Z)) \psi_{N, Z} \right \rangle + \left \langle \abs{x_{N}}^{2} \ \psi_{N, Z}, -\Delta_{N} \psi_{N, Z} \right \rangle \nonumber \\
&\qquad + \left \langle \psi_{N, Z}, \qty(-Z\abs{x_{N}} + \dfrac{1}{N} \sum_{1 \leq i < j \leq N} \dfrac{\abs{x_{i}}^{2} + \abs{x_{j}}^{2}}{\abs{x_{i} - x_{j}}}) \psi_{N, Z} \right \rangle.
\end{align}

\noindent Here, we used the symmetry of $\abs{\psi_{N, Z}}^{2}$ under interchange of particle coordinates, which is true for both fermions and bosons. The first term on the right side of \eqref{eq:eigenvalue_2} is non-negative since

\begin{equation}
\label{eq:eigenvalue_3}
H_{N - 1, Z} \geq E\qty(N - 1, Z) \geq E\qty(N, Z),
\end{equation}

\noindent which follows from the assumption that $E\qty(N, Z)$ is an eigenvalue of $H_{N, Z}$ corresponding to the eigenfunction $\psi_{N, Z} \in H^{1}\qty(\mathds{R}^{3N})$. Hence,

\begin{equation}
\label{eq:eigenvalue_4}
0 \geq \left \langle \abs{x_{N}}^{2} \ \psi_{N, Z}, -\Delta_{N} \psi_{N, Z} \right \rangle + \left \langle \psi_{N, Z}, \qty(-Z + \alpha_{N}\qty(N - 1)) \abs{x_{N}} \psi_{N, Z} \right \rangle,
\end{equation}

\noindent where we have used the definition of $\alpha_{N}$ given in \cite{Nam2012}

\begin{equation}
\label{eq:alpha_definition}
\alpha_{N}:=\inf_{\vb{x} \in \mathds{R}^{3}} \dfrac{\sum_{1 \leq i < j \leq N} \dfrac{\abs{x_{i}}^{2} + \abs{x_{j}}^{2}}{\abs{x_{i} - x_{j}}}}{\qty(N - 1) \sum_{i = 1}^{N} \abs{x_{i}}}.
\end{equation}

\noindent In \eqref{eq:alpha_definition}, $\qty{\alpha_{N}}_{N = 2}^{\infty}$ is an increasing sequence such that $\alpha_{N} \to \beta$ as $N \to \infty$, see \cite{Nam2012}. Let us divide \eqref{eq:eigenvalue_4} by $I = N \left \langle \abs{x_{N}} \psi_{N, Z}, \psi_{N, Z} \right \rangle \geq 0$ to get,

\begin{equation}
\label{eq:eigenvalue_5}
\alpha_{N}\qty(N - 1) \leq Z - \dfrac{\left \langle \abs{x_{N}}^{2} \psi_{N, Z}, -\Delta_{N}\psi_{N, Z} \right \rangle}{\left \langle \abs{x_{N}}\psi_{N, Z}, \psi_{N, Z} \right \rangle},
\end{equation}

\noindent As before, we also use Nam's result,

\begin{equation}
\label{eq:eigenvalue_6}
\Re \left \langle x^{2} f, -\Delta f \right \rangle \geq -\dfrac{3}{4} \left \langle f, f \right \rangle,
\end{equation}

\noindent with $f = \psi_{N, Z}$. From \eqref{eq:eigenvalue_5} and \eqref{eq:eigenvalue_6}, we obtain,

\begin{align}
\label{eq:eigenvalue_7}
\alpha_{N}\qty(N - 1) \leq Z + \dfrac{3}{4} \qty(\left \langle \abs{x_{N}} \psi_{N, Z}, \psi_{N, Z} \right \rangle)^{-1} &\leq Z + \dfrac{3}{4} \left \langle \abs{x_{N}}^{-1} \psi_{N, Z}, \psi_{N, Z} \right \rangle = Z + \dfrac{3}{4} \dfrac{A_{\psi}}{Z},
\end{align}

\noindent where we have used Jensen's inequality from first to second inequality of \eqref{eq:eigenvalue_7}. Taking into account the definition of the Single Particle Density, we can write,

\begin{equation}
\label{eq:eigenvalue_8}
\left \langle \abs{x_{N}}^{-1} \psi_{N, Z}, \psi_{N, Z} \right \rangle = \dfrac{1}{N} \int_{\mathds{R}^{3}} \dfrac{\rho_{\psi}\qty(x_{N})}{\abs{x_{N}}} \dd{x_{N}}.
\end{equation}

\noindent Using the Coulomb Uncertainty Principle again, the definition of $K_{\rho_{\psi}}$ given by \eqref{eq:hoffmann_ostenhof_inequality}, the fact that $\norm{\rho_{\psi}} = N$, and the Virial Theorem in its $N$-particle version, $-E\qty(N, Z) = K_{\psi}$, we get,

\begin{equation}
\label{eq:eigenvalue_9}
\dfrac{A_{\psi}}{Z} = \dfrac{1}{N} \int_{\mathds{R}^{3}} \dfrac{\rho_{\psi}\qty(x_{N})}{\abs{x_{N}}} \dd{x_{N}} \leq K_{\rho_{\psi}}^{1/2} N^{-1/2} \leq \sqrt{-E\qty(N, Z)}N^{-1/2}.
\end{equation}

\noindent Thus,

\begin{equation}
\label{eq:eigenvalue_10}
\alpha_{N}\qty(N - 1) \leq Z + \dfrac{3}{4} \sqrt{-E\qty(N, Z)} N^{-1/2},
\end{equation}

\noindent and using \eqref{eq:estimating_3}, we finally conclude,

\begin{equation}
\label{eq:eigenvalue_11}
\alpha_{N}\qty(N - 1) \leq Z + \dfrac{3}{4}N^{-1/2} \qty(C_{\textrm{IE}}Z^{3} + D\sqrt{\dfrac{1}{2}}N^{4/3}Z)^{1/2}.
\end{equation}

\subsection*{Controlling the square--root term.}
\noindent Before applying the lower bound on $\alpha_{N}\qty(N - 1)$ (see \cite{Nam2012}),

\begin{equation}
\label{eq:errorterm_1}
\beta \geq \alpha_{N}\qty(N - 1) \geq N\qty(\beta - 3\qty(\dfrac{\beta}{6})^{1/3}N^{-2/3}),
\end{equation}

\noindent we observe that \eqref{eq:eigenvalue_11} has a square-root term. Extracting $C_{\textrm{IE}}Z^{3}$ as a prefactor and scaling properly the powers of $Z$ involved, we get

\begin{equation}
\label{eq:errorterm_2}
\alpha_{N}\qty(N - 1) \leq Z + \dfrac{3}{4}Z C_{\textrm{IE}}^{1/2} \qty(\dfrac{Z}{N})^{1/2} \qty(1 + \dfrac{D\sqrt{1/2}}{C_{\textrm{IE}}} \qty(\dfrac{N}{Z})^{4/3} Z^{-2/3})^{1/2}.
\end{equation}

\noindent Using that $(1+x)^{1/2} \leq 1 + x/2$, it follows,

\begin{equation}
\label{eq:errorterm_3}
\alpha_{N}\qty(N - 1) \leq Z\qty(1 + \dfrac{3}{4}C_{\textrm{IE}}^{1/2} \qty(\dfrac{Z}{N})^{1/2}) + \dfrac{3}{8} \dfrac{D}{\sqrt{2 C_{\textrm{IE}}}} \qty(\dfrac{N}{Z})^{5/6} Z^{1/3}.
\end{equation}

\noindent Thus, using \eqref{eq:errorterm_1} in \eqref{eq:errorterm_3}, distributing terms, rearranging $N/Z$ powers, and dividing by $\beta$, it follows that

\begin{equation}
\label{eq:errorterm_4}
N\qty(Z) \leq \dfrac{1}{\beta} \qty(1 + \dfrac{3}{4} C_{\textrm{IE}}^{1/2} \qty(\dfrac{Z}{N})^{1/2})Z + h\qty(N, Z)Z^{1/3}, 
\end{equation}

\noindent with

\begin{equation}
\label{eq:errorterm_5}
h\qty(N, Z):=\dfrac{3}{8\beta} \dfrac{D}{\sqrt{2 C_{\textrm{IE}}}} \qty(\dfrac{N}{Z})^{5/6} + \dfrac{3}{\beta} \qty(\dfrac{\beta}{6})^{1/3} \qty(\dfrac{N}{Z})^{1/3}.
\end{equation}

\subsection*{Final result.} Set $a := \beta^{-1}$, $b:=\dfrac{3}{4\beta}C_{\textrm{IE}}^{1/2}$, $c:=h(N,Z)Z^{-2/3}$, and $x = N(Z)/Z$. Then, \eqref{eq:errorterm_4} reads,

\begin{equation}
	\label{eq:finalterm_1}
	x \leq a + \frac{b}{\sqrt{x}} + c,
\end{equation}

\noindent which in turn implies $x \leq \hat{x}(a+c)$, where $\hat{x}(\mu)$ denotes the unique positive solution of 

\begin{equation}
	\label{eq:finalterm_2}
	\hat{x} = \mu + \frac{b}{\sqrt{\hat{x}}}.
\end{equation}

\noindent To conclude this, we used the fact that 

$$
g(x):=\frac{x}{\mu + \dfrac{b}{\sqrt{x}}}
$$

\noindent is an increasing function in $x$. It is clear, from the definition of $\hat{x}(\mu)$, that $\hat{x}(\mu)$ is an increasing function of $\mu$. Finally, it is simple to show that $\hat{x}(a+c) \leq \hat{x}(a) + c$. In fact, from the definition of $\hat{x}$, we see that,

\begin{equation}
	\label{eq:finalterm_3}
	\hat{x}(a+c) - \hat{x}(a) = c + \dfrac{b}{\sqrt{\hat{x}(a+c)}} - \dfrac{b}{\sqrt{\hat{x}(a)}} \leq c,
\end{equation}

\noindent because $\hat{x}(\mu)$ is increasing in $\mu$. 

\medskip

From \eqref{eq:finalterm_3} and the definitions of $a$, $b$, $c$, and $x$, we get,

\begin{equation}
\label{eq:final_4}
\dfrac{N\qty(Z)}{Z} \leq 1.4811 + h\qty(N, Z)Z^{-2/3}.
\end{equation}

\noindent Here, $1.4811$ is the numerical value of $\hat{x}(a)$, and $h\qty(N, Z)$, the error term, given by \eqref{eq:errorterm_5}. This error term can be estimated by using Lieb's bound \cite{Lieb1984-1} of 1984, $N\qty(Z) < 2Z + 1$, such that $N\qty(Z)/Z < 2 + 1/Z$, for all $Z$,

\begin{equation}
\label{eq:final_5}
h\qty(2 + \dfrac{1}{Z}, Z) = \dfrac{3}{8\beta} \dfrac{D}{\sqrt{2 C_{\textrm{IE}}}} \qty(2 + \dfrac{1}{Z})^{5/6} + \dfrac{3}{\beta} \qty(\dfrac{\beta}{6})^{1/3} \qty(2 + \dfrac{1}{Z})^{1/3}.
\end{equation}

\noindent Using $D \leq 0.4433$, $C_{\textrm{IE}} = 0.1242$, and $\min \beta = 0.8218$, we can compute \eqref{eq:final_5} as an error term for the factor $1.4811Z$. A numerical computation shows that, from $Z = 12$, the upper bound given by Theorem \ref{thm:excess_charge} improves Lieb's bound \cite{GonzalezBrantes2024}. Therefore, using the fact that $h\qty(2 + 1/Z, Z)$ is a decreasing function of $Z$, and that $h\qty(12) \leq 3.1516$, we finally get the upper bound

\begin{equation}
\label{eq:final_6}
N\qty(Z) < 1.4811Z + 3.1516Z^{1/3},
\end{equation}

\noindent for any $Z \geq 12$ which establishes the Theorem \ref{thm:excess_charge}. $\square$

\section*{Acknowledgments}
\noindent \thanks{This work has been supported by Fondecyt Project \# 124--1863 (Chile). We are deeply grateful to Dirk Hundertmark and Marvin Schulz for pointing out errors in a previous version of this manuscript and for many helpful discussions. JMG also thanks ANID (Chile) for their support through a {\it Beca de Doctorado Nacional}, Folio 21212245.}

\end{document}